\documentclass[10pt]{article}
\usepackage{graphicx}
\usepackage{float}
\usepackage{amsmath}
\usepackage{amsfonts}
\usepackage{jheppub}
\usepackage{multirow}
\usepackage{colortbl}
\definecolor{kugray5}{RGB}{224,224,224}
\usepackage{color}

\def\spaa#1.#2.#3{\langle\mskip-1mu{#1}|#2|{#3}\mskip-1mu\rangle}
\def\spbb#1.#2.#3{[\mskip-1mu{#1}|#2|{#3}\mskip-1mu]}
\def\spa#1.#2{\left\langle#1\,#2\right\rangle}
\def\spb#1.#2{\left[#1\,#2\right]}
\def\spab#1.#2.#3{\left\langle#1|#2|#3\right]}
\def\spba#1.#2.#3{\left[#1|#2|#3\right\rangle}

\def\P3b{\bar{P}_3}

\def\g0{\gamma_0}

\def\bentarrow{\:\raisebox{1.3ex}{\rlap{$\vert$}}\!\rightarrow}                 
                                                                    
\def\bothdk#1#2#3#4#5{
\begin{array}{r c l}                                                            
#1 & \rightarrow & #2#3 \\
 & & \:\raisebox{1.3ex}{\rlap{$\vert$}}\raisebox{-0.5ex}{$\vert$} 
\phantom{#2}\!\bentarrow #4 \\
 & & \bentarrow #5                                                              
\end{array}                                                                     
}                                                                               
\newcommand{\cg}{c_\Gamma}

\newcommand{\beq}{\begin{equation}}
\newcommand{\eeq}{\end{equation}}
\newcommand{\beqn}{\begin{eqnarray}}
\newcommand{\eeqn}{\end{eqnarray}}

\newcommand{\nn}{\nonumber}

\newcommand{\slsh}{\rlap{$\;\!\!\not$}}     

\def\vec#1{{\mbox{\boldmath$#1$}}}

\newcommand{\e}{\mbox{$\vec{e}$}}

\title{Single top production in association with a Z boson at the LHC}

\author{
John Campbell, R. Keith Ellis and Raoul R{\"o}ntsch\\
    Fermilab, Batavia, IL 60510, USA\\
    E-mail:
    {\tt johnmc@fnal.gov},
    {\tt ellis@fnal.gov},
    {\tt rontsch@fnal.gov}.}

\preprint{FERMILAB-PUB-13-043-T}

\abstract{We present results for the production of a $Z$ boson
in association with single top at next-to-leading order (NLO), including the decay of the top quark and the $Z$ boson. This electroweak process 
gives rise to the trilepton signature $l^+ l^- l^{\prime\; \pm}$
 + jets + missing energy. We present results for this signature and show that the rate is competitive
 with the contribution of the mixed strong and electroweak
production process, $t\bar{t}Z$. As such it should be observable in 
the full data sample from LHC running at $\sqrt{s}=8$~TeV. 
The single top + $Z$ process is a hitherto unconsidered irreducible background
in searches for flavour changing neutral current decays of the top quark in $t \bar{t}$ production.
For a selection of cuts used at the LHC involving a $b$-tag it is the dominant background.
In an appendix we also briefly discuss
the impact of NLO corrections on the related $tH$ process.}

\keywords{QCD, Phenomenological Models, Hadronic Colliders, LHC}

\begin{document}

\maketitle

\section{Introduction}

After only one year of $8$~TeV running, the LHC has already become a tool
for  detailed studies of the top quark. With an increase to a higher centre-of-mass
energy and anticipated integrated luminosities of up to $3000$~fb$^{-1}$, the
LHC will be able to achieve measurements of unprecedented precision in the top sector.
With the advent of high statistics top physics, it will be possible to study not only the
production of top quark pairs but also processes in which
a vector boson is produced in association with top quarks.

The CMS and ATLAS collaborations have produced first results on $t \bar{t}Z$ and
$t \bar{t}W$ production in recent publications~\cite{CMS-PAS-TOP-12-014,ATLAS-CONF-2012-126}.
The $t \bar{t}W$ process does not depend on the details of the top sector 
since the accompanying $W$ boson is radiated from the initial state quarks. In contrast,
the $t \bar{t}Z$ process directly probes the coupling of the $Z$ boson to the top quark.
Theoretical predictions are available for these processes at the NLO parton
level~\cite{Lazopoulos:2008de,Kardos:2011na,Campbell:2012dh} and in NLO 
calculations matched to a parton shower~\cite{Garzelli:2011is,Garzelli:2012bn}.

\begin{figure}[t]
\begin{center}
\includegraphics[angle=270,width=10cm]{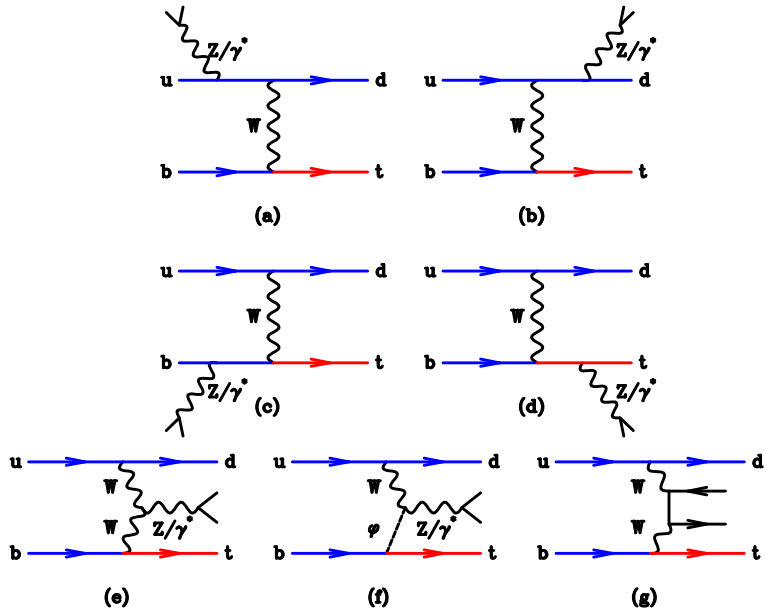}
\caption{Feynman graphs to calculate the lowest order amplitudes.
The wavy line denotes a $W$ or $Z/\gamma^*$ boson.}
\label{fig:STZ}
\end{center}
\end{figure}

In this context it is also interesting to consider the process where an extra 
$Z$ boson is radiated in $t$-channel single top production. 
This predominantly proceeds through the leading order processes,
\beq
u + b \rightarrow d + t + Z \;, \qquad \bar d + b \rightarrow \bar u + t + Z \;,
\label{eq:tZprocess}
\eeq
for the production of a top quark, with smaller contributions from strange- and charm-initiated
reactions. Production of an anti-top quark proceeds through the charge conjugate processes,
\beq
d + \bar b \rightarrow u + \bar t + Z \;, \qquad \bar u + \bar b \rightarrow \bar d + \bar t + Z \;,
\label{eq:tbarZprocess}
\eeq
with a smaller rate at the LHC due to the difference in up- and down-quark parton distribution
functions (pdfs).
The leading order (LO) Feynman diagrams for the first process in Eq.~(\ref{eq:tZprocess}) are shown
in Fig.~\ref{fig:STZ}, including also the non-resonant contribution, diagram $(g)$, that should
be included when considering the charged lepton final state. The $Z$ boson can be radiated from
any of the four quark lines, or from the $W$ boson exchanged in the $t$-channel.
As can be seen from the diagrams, this process is related to hadronic $WZ$ production by crossing.
As a matter of principle, measurement of single top+$Z$ is thus as important as measuring the 
$WZ$ pair cross section, with the added bonus that it depends on 
the coupling of the top quark to the $Z$. In this paper, we present results for the single top + $Z$ process to next-to-leading order (NLO) in QCD\footnote{Next-to-leading order QCD corrections to $tZ$ associated production via the flavor-changing neutral-current
couplings at hadron colliders have been considered in Ref.~\cite{Li:2011ek}.}.

\begin{figure}[t]
\begin{center}
\includegraphics[angle=270,width=0.9\textwidth]{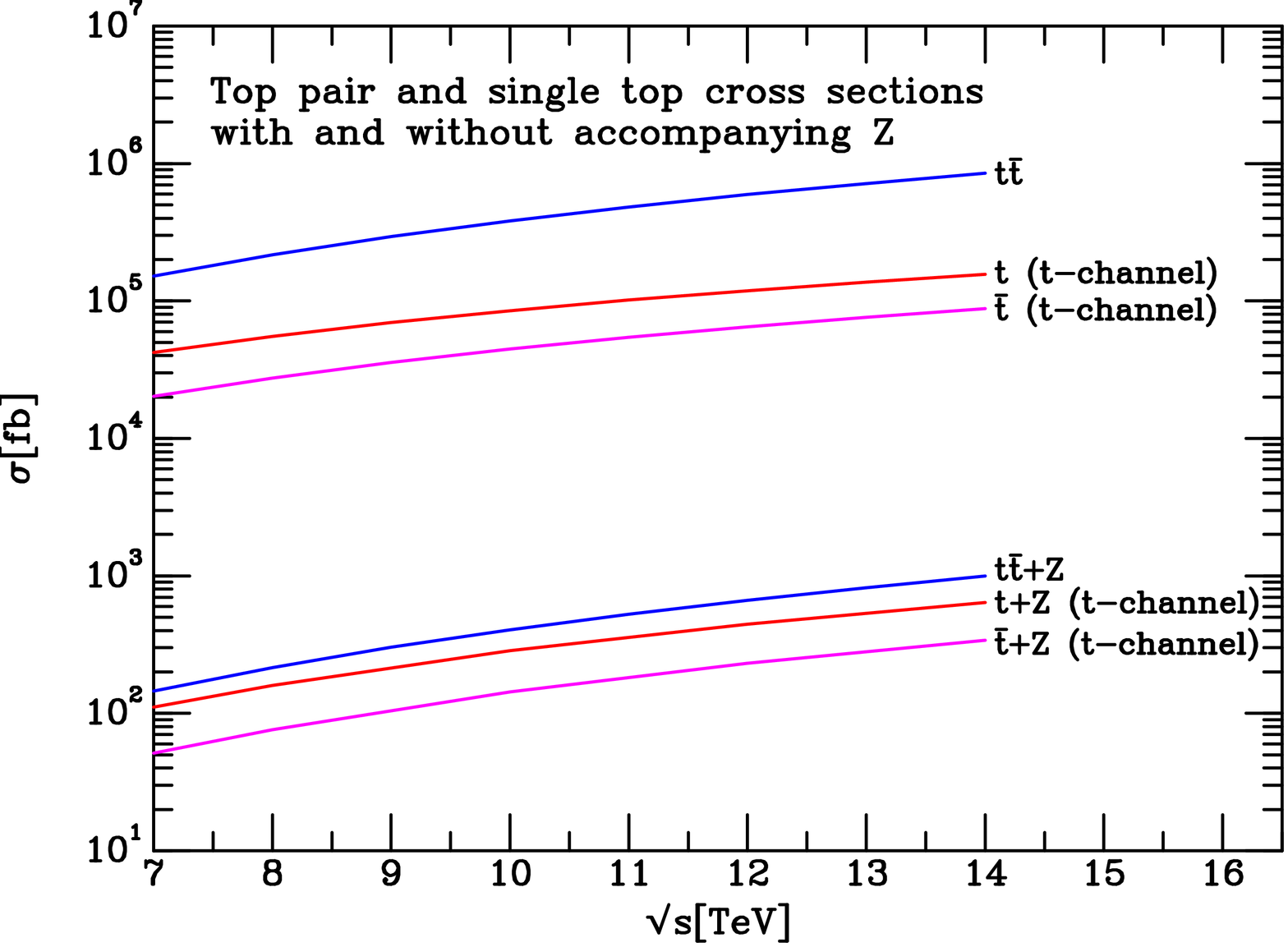}
\caption{NLO inclusive cross sections for single and top quark pair production 
with and without an accompanying Z boson. The NLO $t \bar{t}Z$ cross section is 
estimated from the lowest order result using a $K$-factor of 1.39
and renormalization and factorization scales $\mu = m_t + m_Z/2$~\cite{Kardos:2011na}.}
\label{fig:TplusZ}
\end{center}
\end{figure}

Although the single top~+~$Z$ process is an electroweak one, in contrast to the QCD-induced pair production
mode ($t{\bar t}Z$), it contains fewer particles in the final state and is therefore easier to produce.
Fig.~\ref{fig:TplusZ} shows that any advantage in rate for the top pair production is effectively removed 
once an additional $Z$-boson is required. 
As a result, the single top + $Z$ cross section is about the same size as the $t\bar{t}Z$ one. 
Given the status of current LHC searches for $t{\bar t}V$ production it is interesting
to consider the expected experimental sensitivity to the single top + $Z$ channel. In particular, the
impact of these SM processes should already be present in current trilepton searches, albeit in
regions of lower jet multiplicity. 

In order to properly assess the expected event rates in trilepton searches, in this paper
we will consider the full process (and similarly for the charge conjugate process),
\beq
\bothdk{u+b} {t+}{Z+d}{\mu^- + \mu^+}{\nu +e^+ + b}
\label{eq:tZdk}
\eeq
where the leptonic decay of the top quark is included and we have specified the charged leptons
that are associated with the $Z$ decay. The top quark decay is included using the techniques
described in Refs.~\cite{Kleiss:1988xr,Campbell:2004ch,Campbell:2012uf} and retains all spin
correlations at the expense of requiring the top quark to be treated exactly on-shell. Since
this calculation involves an incoming $b$-quark it is necessarily a five-flavor calculation.

We have also considered the closely-related single top + $H$ process which is of smaller
phenomenological interest in the Standard Model. A brief description of the next-to-leading
order result is given in Appendix~\ref{tHappendix}.

\section{Outline of calculation}
\label{outline}

\subsection{Leading order}
The leading order diagrams for this process are shown in Fig.~\ref{fig:STZ}. It is useful to consider the contribution from
combinations of individual diagrams as follows:
the $Z/\gamma^\star$ attached to the light quark line, $M^{(a,b)}$, 
the $Z/\gamma^\star$ attached to the heavy quark line, $M^{(c,d)}$,
the $Z/\gamma^\star$ attached to the $t$-channel $W$ boson $M^{(e,f)}$,
the non-resonant contribution with the lepton line attached to the $t$-channel exchanged $W$ bosons, $M^{(g)}$.
The computation of the amplitude can be performed in the unitary gauge. 
However a more compact expression is obtained in the Feynman gauge after the inclusion
of an additional contribution representing the propagation of unphysical Higgs fields (represented by $\varphi$ in diagram (f)).
In the latter approach the cancellation of the terms associated with the longitudinal degrees of freedom is built-in.
The explicit form of the leading order amplitudes is given in Appendix~\ref{appendix}.

\subsection{Next-to-leading order}

Next-to-leading order corrections to the single top~+~$Z$ process are computed in a fairly straightforward
manner. 
Virtual corrections to diagrams in which the $Z$ boson is radiated from the $t$-channel
$W$ or in which the lepton pair are produced in a non-resonant manner (c.f. Fig.~\ref{fig:STZ}(e,f) and (g))
consist solely of vertex corrections and are therefore easily computed analytically. For the
remaining diagrams, where the $Z$ boson is radiated from one of the fermion lines, some of the vertex
corrections can be computed in a similar fashion.
However, the virtual amplitude also receives contributions
from box diagrams containing three powers of the loop momentum. These corrections are computed numerically
using a variant of the van Oldenborgh-Vermaseren scheme for the calculation of tensor integrals~\cite{vanOldenborgh:1989wn}.
Scalar integrals are computed using the QCDLoop library~\cite{Ellis:2007qk}.
We have also implemented a version of the usual Passarino-Veltman reduction algorithm~\cite{Passarino:1978jh},
supplemented by special handling of regions of small Gram or Cayley determinants according to the procedure outlined in
Ref.~\cite{Denner:2005nn}.
In our implementation we find that the alternate reduction methods are used to
improve the numerical stability of the calculation in approximately $0.3\%$ of all events.

As a further check, we compare the numerical calculation of the singular contributions to the amplitude
to the known analytic form (after renormalization)~\cite{Catani:2000ef},
\beq
g^2 \cg C_F \left\{\Big(\frac{\mu^2}{s_{16}}\Big)^\epsilon \left[-\frac{2}{\epsilon^2}-\frac{3}{\epsilon}\right]
  +\Big(\frac{\mu^2}{s_{25}}\Big)^\epsilon \left[-\frac{2}{\epsilon^2}-\frac{5}{2 \epsilon}\right]
  +\Big(\frac{\mu^2}{m_t^2}\Big)^\epsilon \left[\frac{1}{\epsilon^2} + \frac{3}{2\epsilon}\right] \right\} \;.
\eeq
where the invariants $s_{25}$ and $s_{16}$ are taken from the momentum assignment in equation \eqref{basicamp}. The overall factor $\cg$ is,
\begin{equation}
\cg = \frac{1}{(4\pi)^{2-\epsilon}}
\frac{\Gamma(1+\epsilon)\Gamma^2(1-\epsilon)}{\Gamma(1-2\epsilon)}\ .
\end{equation}
We find that less than $0.02\%$ of all events fail this consistency check and are discarded. Moreover, these points
lie in extreme phase space regions that contribute little to total cross sections. When realistic experimental
cuts are applied the proportion of numerically unstable points removed from the calculation drops
by a factor of about four.

The calculation is performed in the four-dimensional helicity (FDH) scheme~\cite{Bern:2002zk}.
The mass renormalization is fixed by the condition that
the inverse propagator vanish on-shell. In the FDH scheme we have,
\begin{equation} \label{massrenormalization}
Z_m = 1- \cg g^2 C_F
 \Bigg[ \frac{3}{\epsilon} +3 \ln\left(\frac{\mu^2}{m^2}\right)+5\Bigg] +\ldots \;,
\end{equation}
and the wave function renormalization is,
\begin{equation} \label{wavefunction}
Z_Q= 1-g^2 \cg C_F \Bigg[ \frac{3}{\epsilon}+3 \ln\left(\frac{\mu^2}{m^2}\right)+5\Bigg] +\ldots \; .
\end{equation}
The coupling of the scalar $\varphi$ to the quark field, proportional to the top mass, must also be renormalized
in the same way.

The top quark decay is included using the method of Ref.~\cite{Campbell:2012uf}. We have included only
the leading order amplitude for the decay since the rate for this process is already very small.

\section{Results}

\begin{table}
\begin{center}
\begin{tabular}{|l|l||l|l|}
\hline
$m_W$ & $80.398$~GeV & $\Gamma_W$ & $2.1054$~GeV \\
$m_Z$ & $91.1876$~GeV & $\Gamma_Z$ & $2.4952$~GeV \\
$m_t$ & $173.2$~GeV  & $G_F$ & $1.116639\times 10^{-5}$ \\
$\alpha_S^{\rm LO}(m_Z)$ & 0.130 & $\alpha_S^{\rm NLO}(m_Z)$ & 0.118 \\
\hline
\end{tabular}
\caption{Input parameters used for the phenomenological results.
The two values of $\alpha_S(m_Z)$ correspond to the choices made in the
CTEQ6L1 and CTEQ6M pdf sets, used at LO and NLO respectively.}
\label{input}
\end{center}
\end{table}
For the results that we present in this paper, we have used the parameters
listed in Table~\ref{input}. From these, the
Weinberg angle is fixed by the tree-level relation,
\begin{equation}
\sin^2 \theta_W = 1-\frac{m_W^2}{m_Z^2} \;,
\end{equation}
which ensures that the amplitudes are gauge invariant. 
Since our calculation is performed in the five-flavour scheme, with an initial
state massless $b$-quark, we also set $m_b=0$ in the decay of the top quark.
For simplicity we work in the framework of a unit CKM matrix.
The parton distributions employed are the CTEQ6L1 set (used at LO) and
CTEQ6M set (used at NLO) taken from ref.~\cite{Pumplin:2002vw}. The renormalization
and factorization scales, denoted by $\mu_R$ and $\mu_F$ respectively, are taken to be the same
for our standard scale choice, $\mu_R = \mu_F = m_t$.

\begin{figure}[ht]
\begin{center}
\includegraphics[angle=0,width=10cm]{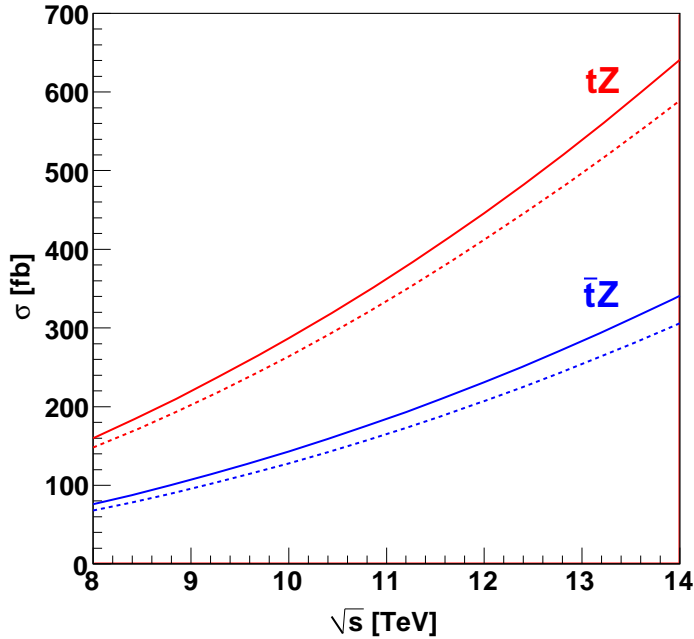}
\caption{Cross sections for $tZ$ and $\bar t Z$ production as a function of $\sqrt s$. The
leading order predictions are shown as dashed lines and the next-to-leading order solid lines.}
\label{tZsqrts}
\end{center}
\end{figure}
With these parameters, the total cross sections for $tZ$ and $\bar t Z$ production as a
function of the LHC operating energy $\sqrt{s}$ are shown in Figure~\ref{tZsqrts}.
Although the leading order process contains a quark, the $t$-channel
exchange of the $W$ boson means that the amplitude does not contain a collinear singularity
and thus that the inclusive cross section is well-defined. The cross section
for $\bar t Z$ production is approximately half the corresponding $t Z$ rate, a reflection
of the corresponding parton distribution function ratio, $f_d(x)/f_u(x) \approx 0.5$ 
at values of $x$ typical of those relevant
for this process, $x \geq (m_t+m_Z)/\sqrt{s} \approx 0.02 - 0.03$. The NLO corrections take a
similar form for both processes, resulting in an increase in the cross section predictions of the
order of $10\%$. Finally, we see that although the cross sections are only of the order of
a few hundred femtobarns at $\sqrt{s}=8$~TeV, these processes have a combined cross section
that is approximately a picobarn at $\sqrt{s}=14$~TeV.

\begin{figure}[ht]
\begin{center}
\includegraphics[angle=0,width=10cm]{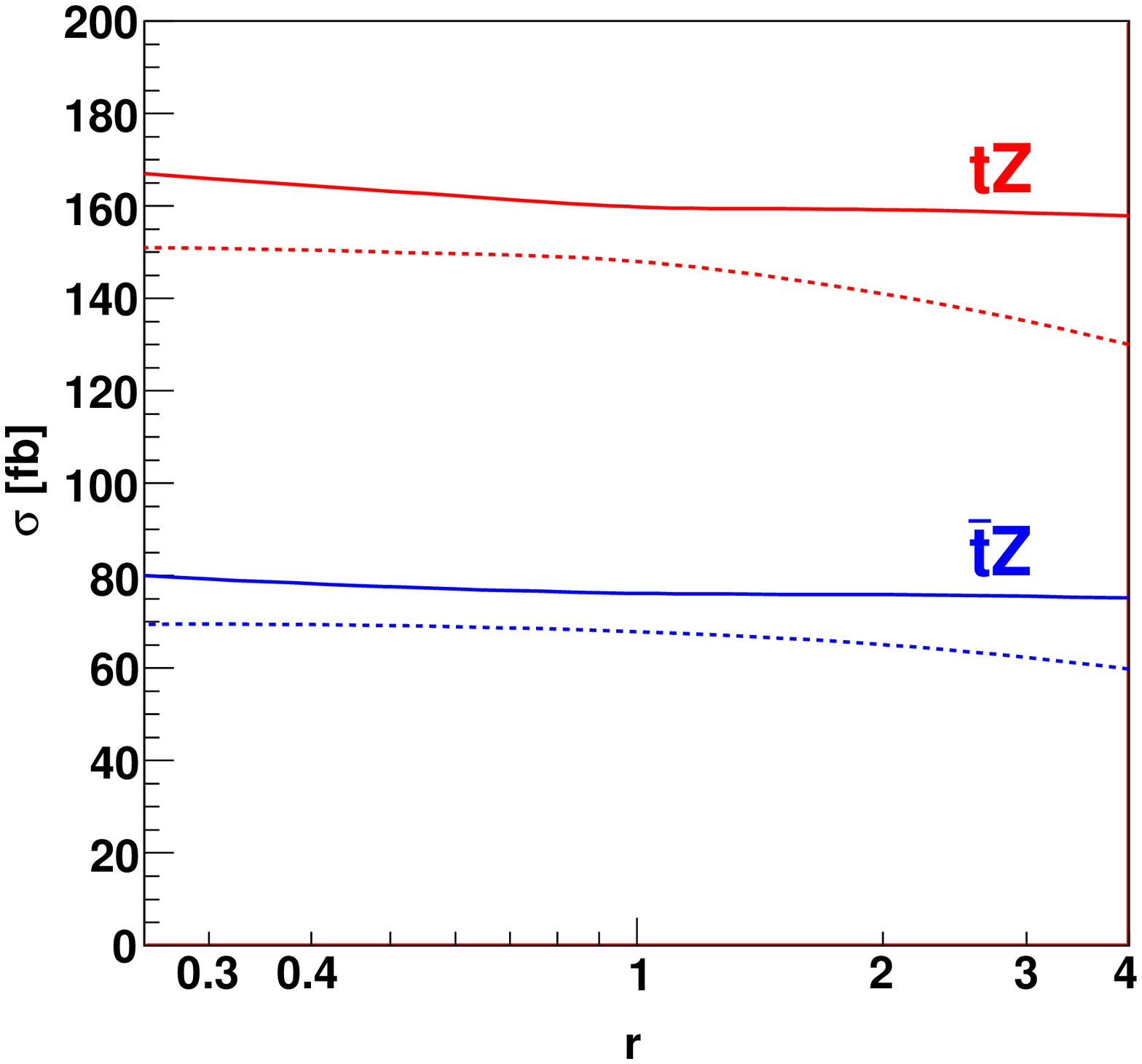}
\caption{Scale dependence of single top + $Z$ cross sections at $8$~TeV.
The renormalization and factorization scales are varied in opposite directions according
to $\mu_R = r \, m_t$, $\mu_F = m_t/r$.}
\label{tZmudep}
\end{center}
\end{figure}
To investigate the scale dependence of this process we focus on the centre-of-mass energy $\sqrt s=8$~TeV.
Since the tree level process does not contain a strong coupling the resulting cross section
only depends on the factorization scale, but at next-to-leading order the renormalization scale
enters for the first time. We find that varying both scales together in the same direction
leads to an accidental cancellation and therefore an artificially small estimate of the scale dependence.
We therefore choose to vary them in opposite directions, $\mu_R = r \, m_t$, $\mu_F = m_t/r$
with $r \in [1/4, 4]$.  The results are shown in Figure~\ref{tZmudep}, where one can see
that the overall scale dependence is still very weak. Even over such a large scale range
the largest deviation from the central value is less than six percent.

Before turning to less inclusive cases, we summarize our findings by presenting predictions
for LO and NLO cross sections at $\sqrt{s}=8$~TeV. For the NLO prediction it is useful to consider
the theoretical uncertainty that should be attributed to the calculation.
In addition to the scale dependence uncertainty,
based on the variation of $r$ over the full range as described above, we also consider the
effect of uncertainties in the extraction of the pdfs. By using the additional uncertainty sets provided
in the CTEQ6 distribution,  we find that this uncertainty is at the level of $7\%$.
We thus find,
\beqn
 & \sigma_{LO}(tZ) = 148~{\rm fb} \;, \qquad &
   \sigma_{NLO}(tZ) = 160^{+7}_{-2}~{\rm (scale)}{}^{+11}_{-11}~{\rm (pdf)}~{\rm fb} \;, \\
 & \sigma_{LO}(\bar tZ) = ~68~{\rm fb} \;, \qquad &
   \sigma_{NLO}(\bar tZ) =  ~76^{+4}_{-1}~{\rm (scale)}{}^{+5}_{-5}~{\rm (pdf)}~{\rm fb} \;,
\eeqn
Combining the two sources of error, the single top~+~$Z$ cross section is thus predicted with
a total uncertainty of just over $10\%$.

\subsection{Comparison of rates for $tZ$, $\bar{t}Z$ and $t\bar{t}Z$}

As discussed in the introduction, the cross-section for $t \bar t Z$
production is comparable to that for the sum of
$tZ$ and $\bar t Z$ production. Referring to equation \eqref{eq:tZdk}, the signature for $tZ$ production is three charged leptons, missing
energy (which can be reconstructed up to the usual two-fold ambiguity) and jets. One of the jets may be $b$-tagged, although we
ignore that possibility in this section.  In the top-pair production
scenario, the subsequent semi-leptonic decay of one top and the
hadronic decay of the other, together with the leptonic decay of the
$Z$-boson, gives rise to the same signature of three charged leptons, missing
energy and jets. If some of the jets go undetected, then the question
arises as to whether it is possible to disentangle these two production processes.

In order to answer this question, we calculate jet-binned
cross-sections for four processes,
\begin{eqnarray}
& {(a)} \; t(\to \nu_e \e^+ b)Z \;, \quad \qquad &
   {(c)} \; t(\to \nu_e e^+ b) \bar{t} (\to q\bar{q} {\bar b}) Z \;, \nonumber \\
&  {(b)} \; \bar{t}(\to e^- \bar{\nu}_{e} \bar{b}) Z \;, \quad \qquad &
   {(d)} \; t(\to q \bar{q} b) \bar{t} (\to e^- \bar{\nu}_{e} \bar{b})Z \;,
\end{eqnarray}
with the decay $Z \to \mu^-\mu^+$ understood in each case.
We perform our comparison at the $\sqrt{s}=14$ TeV LHC. The scale $\mu=m_t$ is used for the $tZ$ and $\bar{t}Z$ calculations, and $\mu=m_t+m_Z/2$ for $t\bar{t}Z$,
following refs. \cite{Lazopoulos:2008de,Kardos:2011na}. We will make use of three sets of kinematic cuts. The first, which we refer to as ``standard cuts'',
requires that the momenta of the leptons, jets and missing energy are each greater than $20$~GeV, and that the pseudorapidity of the leptons and jets are
constrained by $|\eta_l| < 2.5$ and $|\eta_j| < 3.5$. Jets are constructed with the anti-k$_t$ algorithm using $\Delta R=0.4$. The second set of cuts require a
more central jet, $|\eta_j| < 2.0$, but are otherwise the same. We shall refer to these cuts as ``$|\eta_j| < 2.0$'' cuts. The third set of cuts is identical to
the standard cuts, but the jets are constructed using  $\Delta R=0.7$. This is referred to as the ``$\Delta R=0.7$'' setup.

\begin{figure}[ht]
\begin{center}
\includegraphics[scale=0.6]{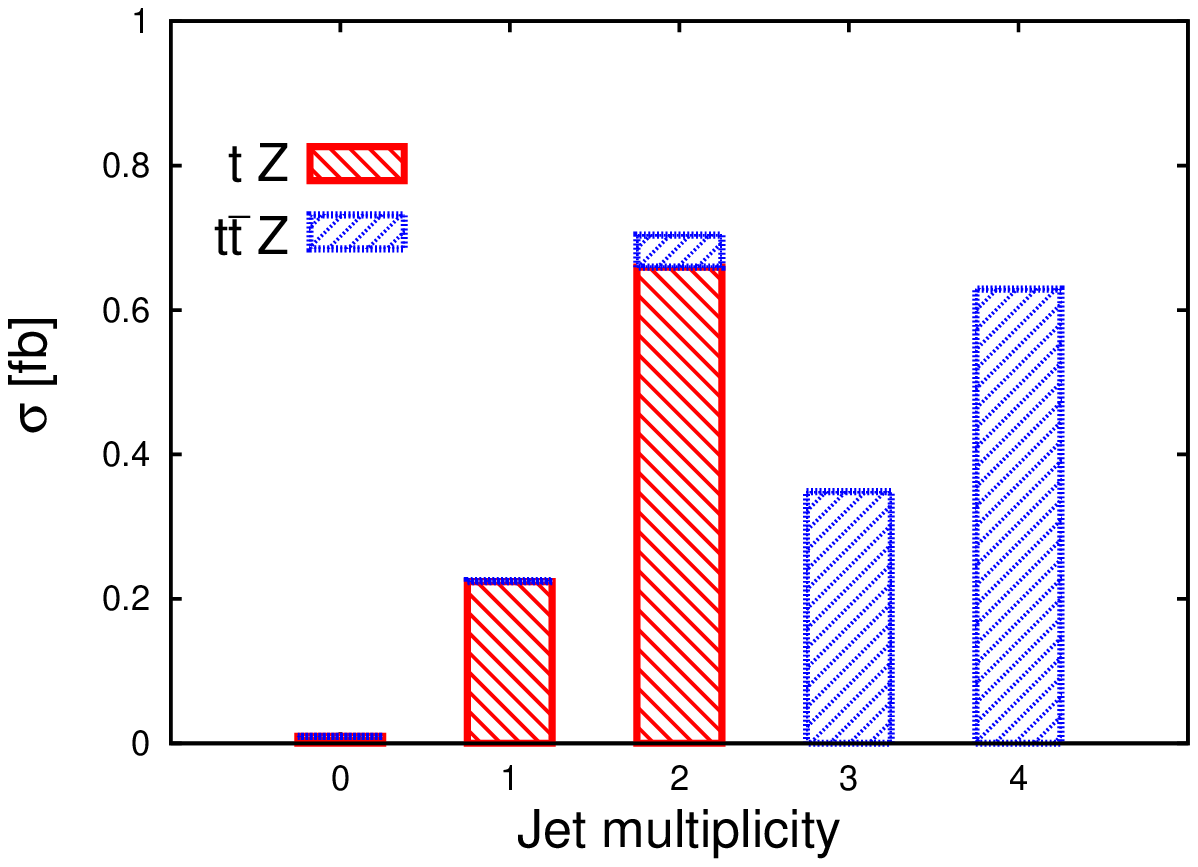}
\includegraphics[scale=0.6]{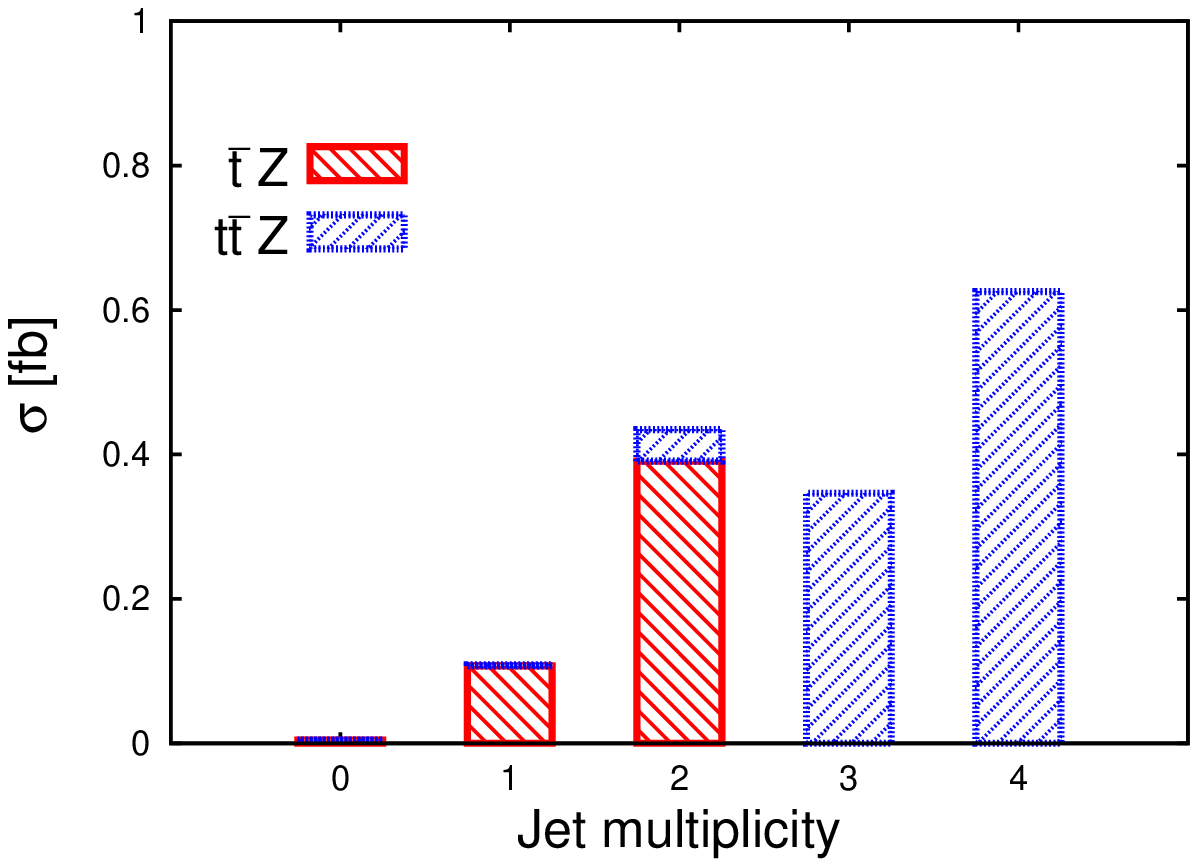} \\
\includegraphics[scale=0.6]{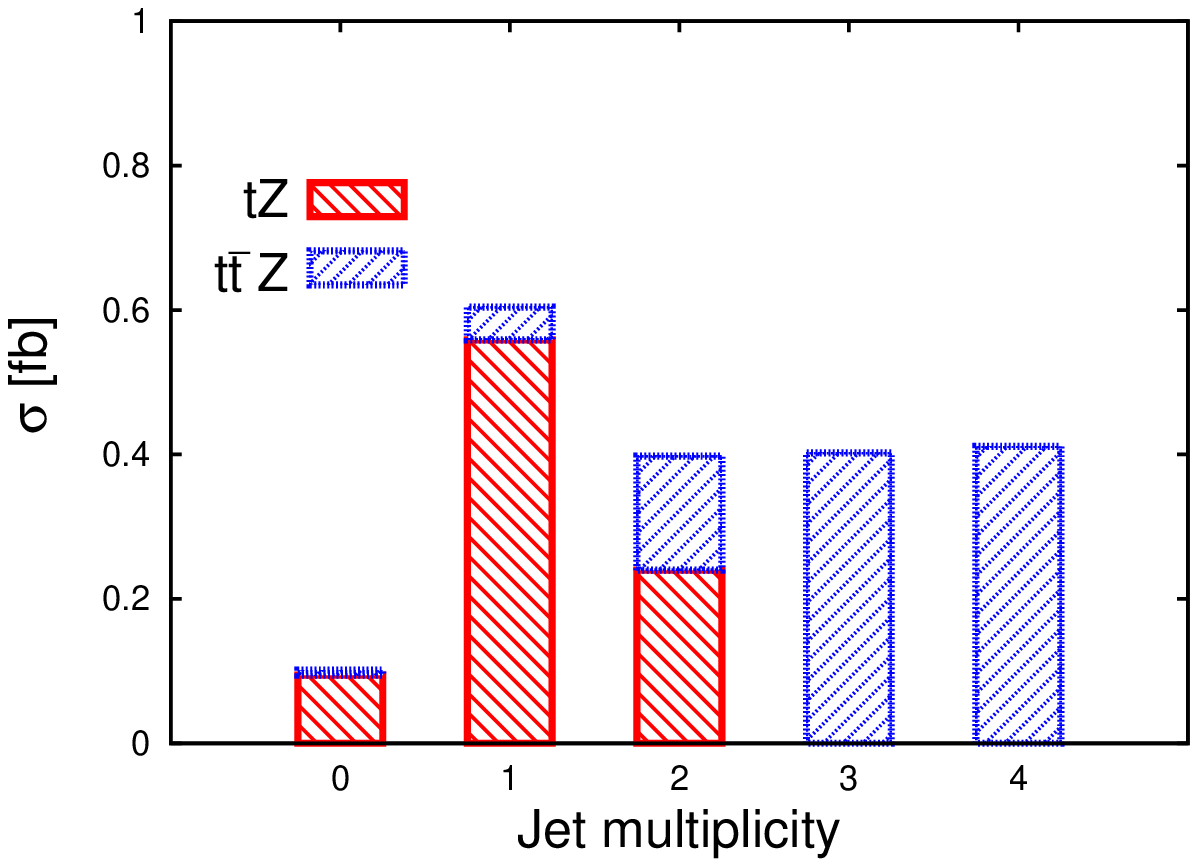}
\includegraphics[scale=0.6]{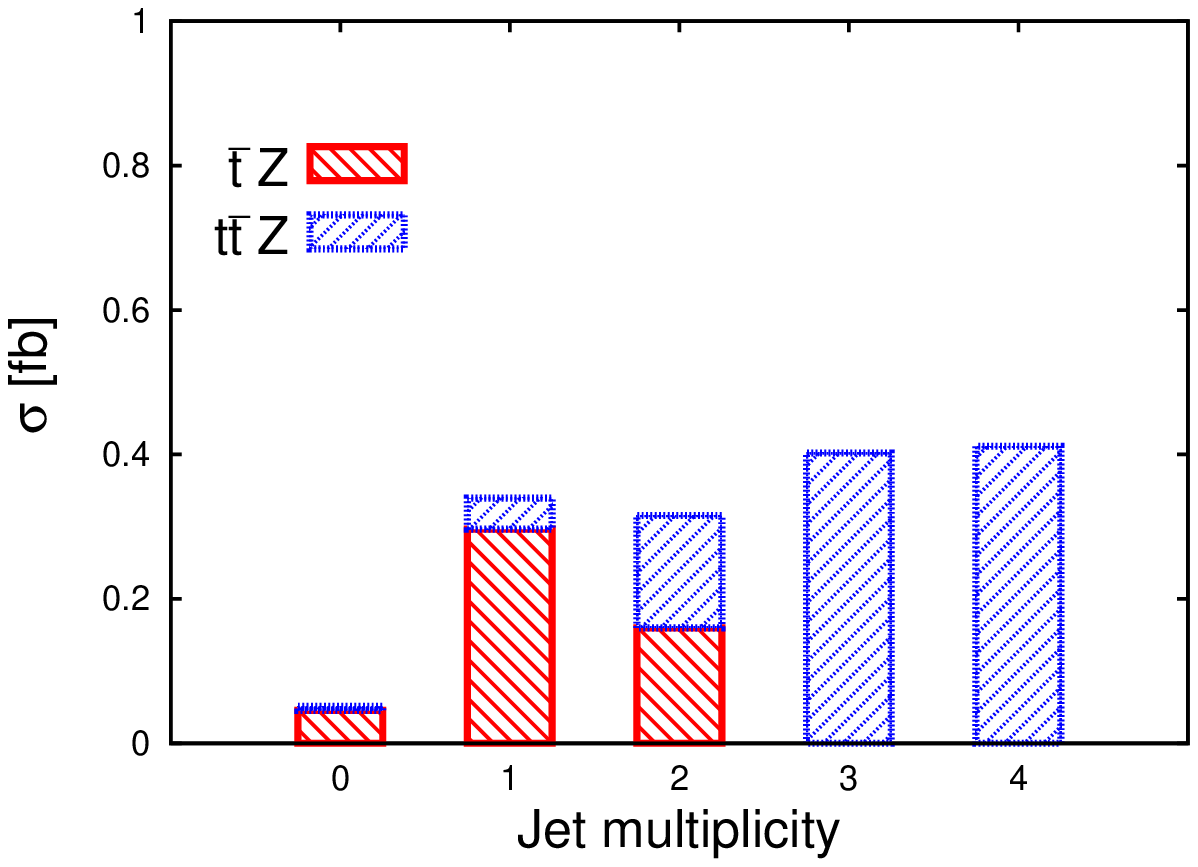} \\
\includegraphics[scale=0.6]{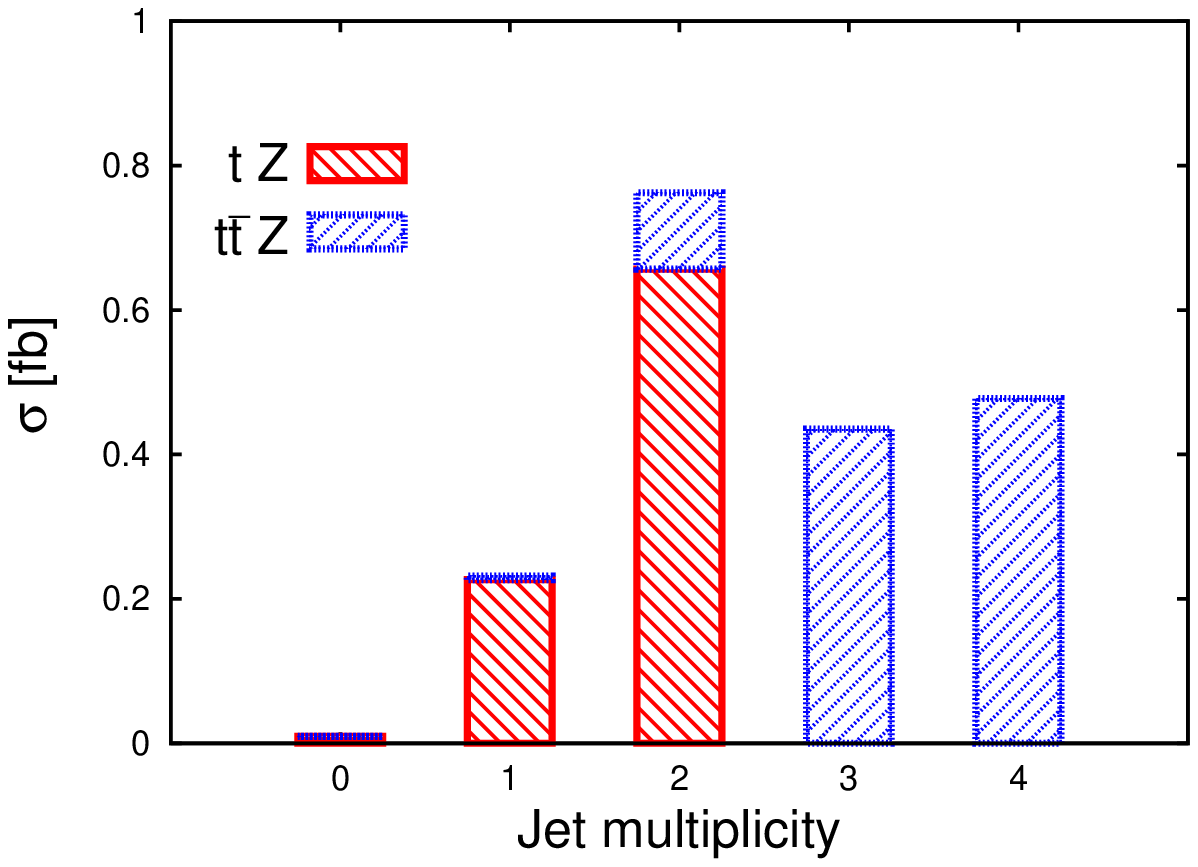}
\includegraphics[scale=0.6]{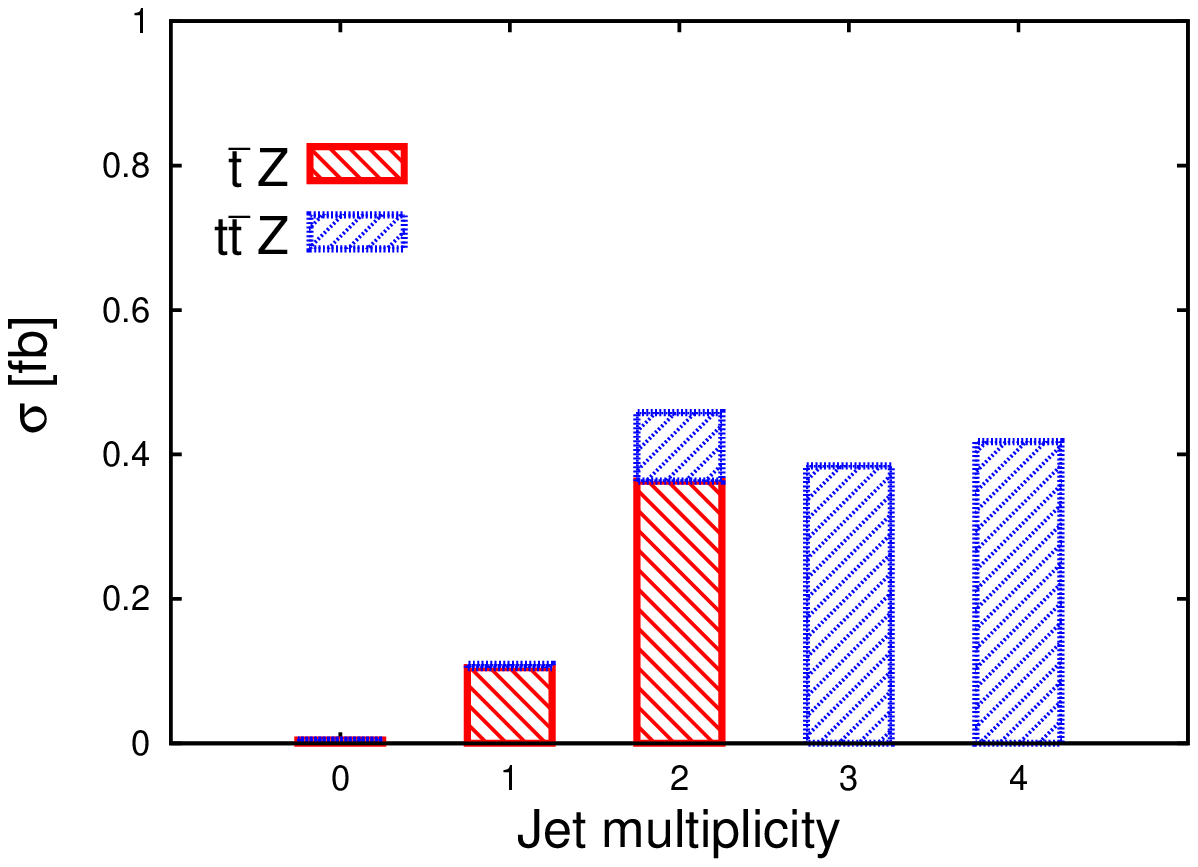} 
\caption{Comparison of jet-binned cross-sections calculated at LO at $\sqrt{s}=14$ TeV.
The left-hand plots show $tZ$ production and $t\bar{t}Z$ production with the subsequent semi-leptonic decay of the top,
resulting in a final state of $\mu^-\mu^+e^+$. The right-hand plots show $\bar{t}Z$ and $t\bar{t}Z$ production with the
subsequent decay of the $\bar{t}$, with a final state of $\mu^-\mu^+e^-$. The first row corresponds to the standard cuts described in the text, the second row uses the $|\eta_j| < 2.0$ cuts, and the final row has the $\Delta R=0.7$ setup. The scale $\mu=m_t$ is used for $tZ$ and $\bar{t}Z$, and $\mu=m_t+m_Z/2$ is used for $t\bar{t}Z$.}
\label{fig:ttZvstZ_LO}
\end{center}
\end{figure}
The comparisons are shown for the LO results in figure~\ref{fig:ttZvstZ_LO}.
The figures on the left are for processes \textit{(a)} and \textit{(c)}, which result in a final state signature with two positively charged leptons; the right-hand figures show processes \textit{(b)} and \textit{(d)}, for which the signature includes two negatively charged leptons. Of
course, the results for the $t\bar{t}Z$ process are the same
irrespective of which top decays hadronically, whereas the $tZ$
cross-sections are a factor of approximately two greater than those
for $\bar{t}Z$, as indicated in figure \ref{tZsqrts}.
This feature suggests a way of distinguishing between the single top + $Z$ and $t\bar{t}Z$ processes, by, for example, considering the
asymmetry between $l^+l^-l'^{+}$ and $l^+l^-l'^{-}$ production. This method would rely on a stringent rejection of backgrounds,
some of which would display a similar asymmetry.

The first row of figure~\ref{fig:ttZvstZ_LO} corresponds to the standard set of cuts. It is seen that most of the jets in $tZ$ production are able
to pass these cuts, so that the two-jet bin dominates the total cross-section. By contrast, the $t\bar{t}Z$ process has a small cross-section in the
two-jet bin and a negligible contribution to the one-jet bin. 

The effect of lowering the cut on the jet pseudorapidity to $|\eta_j| < 2.0$ is shown in the second row. Since one of the jets in $tZ$ production is
usually quite forward, with the other one central, it is unsurprising to see that the one-jet bin is dominant for $tZ$ production. It is also evident
that the stricter jet cut has shifted some of the $t\bar{t}Z$ events to the lower jet bins, with the result that the two-jet bin contains a significant
proportion of events originating from this process. 

The third row shows the results using the $\Delta R=0.7$ setup. This has little
effect on the jets originating from $tZ$ production: since one is
forward and the other one central, there is little opportunity for
these to be clustered into one jet. The effect is more pronounced for
$t\bar{t}Z$, enhancing the cross-section in the two-jet bin.

The effect of NLO corrections to the $tZ + \bar{t}Z$ cross-sections are shown in table \ref{tab:tZNLO}. The total cross-section shows a slight increase from $\sigma_{LO}=1.4$ fb at LO to $\sigma_{NLO}=1.5$ fb at NLO. However, looking at the standard cuts, it is clear that this increase is not uniform over
the jet bins. The three-jet bin contributes around half the total cross-section, indicating that the additional radiated gluon is usually quite hard.
This has the effect of migrating events from one jet bin to the next, with the result that the cross-sections in the zero-, one- and two-jet bins
\textit{decrease} due to the NLO corrections. This holds true when a larger jet is used,  $\Delta R =0.7$, although the two-jet bin is larger and the
three-jet bin smaller than with the standard cuts. This is because of the increased likelihood of clustering the radiated gluon with one of the LO
partons, leaving two jets. When the  $|\eta_j| < 2.0$ cuts are used, the NLO corrections decrease the one-jet bin and increase the two-jet bin. The
three-jet bin is much smaller than for the standard set of cuts. 
\begin{table}
\begin{center}
\begin{tabular}{|c|c|c|c|c|c|}
\hline
\multicolumn{2}{|c|}{Jet multiplicity} & 0 & 1 & 2 & 3 \\
\hline
\multirow{2}{*}{Standard cuts} & LO  & 0.014 & 0.331 & 1.05 & - \\
                               & NLO & 0.011 & 0.237  & 0.585 & 0.693 \\
\hline
\multirow{2}{*}{$|\eta_j| < 2$} & LO  & 0.140 & 0.856  & 0.400 & - \\
                               & NLO & 0.115 & 0.669  & 0.531 & 0.211 \\
\hline
\multirow{2}{*}{$\Delta R =0.7$} & LO  & 0.014 & 0.336  & 1.05 & - \\
                               & NLO & 0.010 & 0.241  & 0.661 & 0.614 \\
\hline
\end{tabular}
\end{center}
\caption{Jet-binned LO and NLO cross-sections (in fb) for $tZ + \bar{t}Z$ production at the $\sqrt{s}=14$~TeV LHC, for the three sets of
cuts described in the text.}
\label{tab:tZNLO}
\end{table}

 The NLO corrections indicate that distinguishing between $tZ$ and $t\bar{t}Z$ production may be more difficult than a LO calculation leads
one to expect. The NLO corrections deplete the $tZ$ cross-sections in the bins where they are dominant over the $t\bar{t}Z$ cross-sections,
and result in comparable cross-sections in the three-jet bin, which only received contributions from $t\bar{t}Z$ at LO. Nor is this the
final story. A more realistic calculation of the jet-binned cross-sections would take parton showering into account. These effects can have
a significant impact on exclusive observables. It should also be borne in mind that NLO corrections and/or parton showering effects may
modify the $t\bar{t}Z$ results. Ideally, a comparison would be performed after calculating both processes to NLO in QCD, and then
interfacing them with a parton showering program that preserves the NLO accuracy.

\subsection{Single top + $Z$ as a background in non-standard top decay searches}

The top quark decays primarily via a $W$ boson, $t \to Wq$, with a
bottom quark being the most likely decay product and the presence of
strange or down quarks suppressed by the off-diagonal CKM
elements. In the standard model, decays through a flavor-changing neutral current (FCNC) are
loop-suppressed, yielding a very small branching ratio $\mathcal{B}(t\to Zq)
< 10^{-12}$~\cite{Glover:2004cy}. Therefore, the
observation of such a decay would be indicative of New Physics.
Searches for FCNC decays in $t\bar{t}$ production were conducted by both CDF~\cite{Aaltonen:2008ac} and D0~\cite{Abazov:2011qf}. Currently,
the best constraints come from  $t\bar{t}$ production at the LHC: ATLAS constrains the branching ratio $\mathcal{B}(t \to Zq) < 0.73\%$ 
with $2.1$~fb$^{-1}$ of data at $\sqrt{s}=7$ TeV~\cite{Aad:2012ij}, while CMS constrains $\mathcal{B}(t \to Zq) <0.24\%$ with
$5.0$~fb$^{-1}$ of data at the same energy~\cite{:2012sd}.

As the second top is taken to decay through the Standard Model mode $t \to Wb$, the signature of these events (with leptonic decays of both
the $W$- and $Z$-bosons) is three charged leptons, missing energy from a neutrino (whose longitudinal momentum is reconstructible, up to the
usual two-fold ambiguity), and two or more jets, one of which can be $b$-tagged. The same signature is expected in $tZ$ and $\bar{t}Z$
production. However, neither the ATLAS~\cite{Aad:2012ij} nor the CMS~\cite{:2012sd} analysis take this background into account. The purpose
of this section is to look at the role of $tZ$ and $\bar{t}Z$ production as a background to FCNC top decays.

We consider decays of the $W$- and $Z$ bosons into different flavored
leptons, $Z \to  \mu^- \mu^+$ and $ W \to \nu_{e} e$, and impose a set
of cuts similar to those used in the CMS analysis\footnote{The cuts used by CMS are slightly more complicated, since they take into account various detector effects.}:
\begin{itemize}
\item Leptons are required to have transverse momentum $p_{T,l} > 20$ GeV and pseudorapidity $|\eta_l| < 2.5$.
\item The missing transverse momentum is constrained by $p_{T,\mathrm{miss}}>30$ GeV.
\item Jets are defined with the anti-$k_T$ algorithm with $\Delta R =0.5$, 
and are required to have $p_{T,j} > 30$ GeV and $|\eta_j| < 2.4$, 
and to be separated from any lepton by $\Delta R_{jl} > 0.4$.
\item The same-flavor dilepton pair is required to have mass $60~ \mathrm{GeV} < m_{ll} < 120~ \mathrm{GeV}$. 
This pair is taken as originating from the $Z$-boson, 
with the remaining lepton originating from the $W$-boson.
\item Each lepton is required to be isolated. In particular, the ratio
of the sum of the transverse energies and momenta of all objects
(leptons and jets) within $\Delta R = 0.3$ of the lepton to the lepton's transverse momentum must be less than 0.125 for leptons
originating from the $Z$-boson, and less than 0.1 for the lepton
originating from the $W$-boson:
\begin{equation*} 
\frac{\sum_{\Delta R_W <0.3} (E_T+p_T)}{p_{T,l}} <  0.1; \hspace{0.4in}
\frac{\sum_{\Delta R_Z <0.3} (E_T+p_T)}{p_{T,l}} <  0.125 
\end{equation*}
(for our purposes, we set $E_T=p_T$).
\end{itemize}

In addition to the above cuts, CMS uses two further sets of cuts, called ``$S_T$'' cuts and ``$b$-tag'' cuts. In the case of the former, the following cuts are applied:
\begin{itemize}
\item At least two jets are required, with the transverse momentum cut as above.
\item The total transverse momentum $S_T = \sum_j p_{T,j} + \sum_l p_{T,l} + p_{T,\mathrm{miss}} > 250$~GeV.
\item  The masses of the $Zj$ and $Wb$-system are constrained to be between $100$~GeV and $250$~GeV.
\end{itemize}

The  ``$b$-tag'' cuts are:
\begin{itemize}
\item At least two jets are required, one of which is $b$-tagged.
\item The masses of the $Zj$- and $Wb$-systems are constrained to be close to the top mass: $| m_{Zj} -m_t| < 25$~GeV and $|m_{Wb}-m_t| < 35$~GeV.
\end{itemize}

The LO and NLO cross-sections for $tZ$ and $\bar{t}Z$ production are
shown at the $\sqrt{s}=7$~TeV LHC in table \ref{tab:fcnc_sf}. There is
a negligible change when the three charged leptons have the same
flavor. We note that the NLO corrections have a substantial effect
on the cross-sections, with a $K$-factor of around 1.5 for the $S_T$
cuts and 1.7 when the $b$-tagging cuts are used. This is because the
additional jet from the real radiation helps to satisfy the jet cuts.
The scale uncertainty is larger than discussed previously for the inclusive production, and we estimate these uncertainties by varying both the factorization
and renormalization scales in the same direction, between $m_t/2$ and $2m_t$. This gives a scale uncertainty of around 5-7\%. The pdf uncertainty is not taken
into account, but is expected to be similar in magnitude.

The dominant background in the CMS analysis comes from $WZjj$ production, with leptonic decay of the weak bosons. Imposing the
$S_T$ cuts we calculate this cross-section to be 0.91 fb
at LO, with a scale uncertainty of around 25\%. Multiplying by a factor of four to include all leptonic final states $eee, ee\mu, \mu\mu
e,\mu\mu\mu$, we find in a sample of $5.0 \mathrm{fb}^{-1}$ that this corresponds to $0.91 \times 4 \times 5 = 18.2$~events.
This is consistent with the CMS calculation of $13.6 \pm 2.6$ $WZjj$ events. 

We can convert the cross sections of table~\ref{tab:fcnc_sf} into event rates to compare
with the CMS study in similar fashion. This implies that 1.6 events should be seen for the $tZ +\bar{t}Z$ background when the
$S_T$ cuts are used. This is a small but not negligible increase on the 16.2 overall background
events that are expected. However, when the $b$-tag cuts are used, the overall CMS background
estimation drops significantly to 0.83 events, due to a more stringent cut on the mass-window of the
weak boson-jet system, and the requirement of a $b$-tag. Since our implementation of $tZ$ production
constrains the $Wb$-system to the top mass and guarantees the presence of a $b$-jet, the effect of
these cuts is far less severe, and we expect 0.74 events coming from the $tZ +\bar{t}Z$ background
with this set of cuts. At present, the best constraint on the FCNC branching ratio is found using
the $S_T$ cuts. However, it is possible that this situation could be changed once the dominant single top + $Z$ contribution to the backgrounds with $b$-tag cuts is included.

\renewcommand{\baselinestretch}{1.4}
\begin{table}
\begin{center}
\begin{tabular}{|c|c|c|c|}
\hline
& & $S_T$ cuts & $b$-tag cuts \\
\hline
\multirow{2}{*}{$Ztj$} & $\sigma_{LO}$ & $33.3(1)^{+1.2}_{-2.0}$ & $14.3(1)^{+0.6}_{-0.8}$ \\
                      & $\sigma_{NLO}$ & $52.0(1)^{-1.6}_{+2.8}$ & $24.5(1)^{-0.9}_{+1.5}$\\
\hline
\multirow{2}{*}{$Z\bar{t}j$} & $\sigma_{LO}$ & $17.5(1)^{+0.6}_{-1.0}$ &~$7.71(1)^{+0.26}_{-0.46}$\\
                      & $\sigma_{NLO}$       & $26.2(1)^{-0.7}_{+1.1}$ & $12.5(1)^{-0.4}_{+0.8}$\\
\hline
\end{tabular}
\end{center}
\renewcommand{\baselinestretch}{1}
\caption{Leading- and next-to-leading order cross-sections (in ab) for $ Z(\to \mu^- \mu^+) t(\to \nu_{e} e b) j$ using the two classes of cuts used in the CMS searches for FCNC in top decays. The cross-sections are evaluated at a scale $\mu=m_t$, with the integration error in the last digit in parentheses. The effect of using a scale choice of $\mu=m_t/2$ and $\mu=2m_t$ are shown as subscripts and superscripts respectively.}
\label{tab:fcnc_sf}
\end{table}
\renewcommand{\baselinestretch}{1}

\section{Conclusions}
We have calculated the production cross-section of single top + $Z$-boson to NLO in QCD, including the leptonic decays of the top quark.  We
have demonstrated that this process is competitive  in rate with the mixed strong and electroweak $t \bar{t}Z$ process. As such, it should
be observable in recorded data from the LHC, despite being subject to  a considerable reducible background from $W^\pm Z $+2 jet processes.
Given this, the potential to constrain the top-$Z$ boson coupling through the $tZ$ process should be investigated further. Moreover, we have
shown that the use of jet-binned cross-sections may be helpful in distinguishing this process from the  $t \bar{t}Z$ process, although this
requires further effort on the theoretical front to determine the effects of parton showering for this observable.
In addition, this process constitutes an irreducible and potentially dominant background in searches for
flavour changing neutral current decays in $t \bar{t}$ production, which is not taken into account in current searches. It will be challenging to remove because, like the  signal, it contains
a real top quark. Code for this phenomenological interesting process,
as well as the related $tH$ process, is included in MCFM v6.6.

\section*{Acknowledgments}
We gratefully acknowledge useful conversations with Kirill Melnikov
and Giulia Zanderighi. We also thank Fabio Maltoni for pointing out
an error in the calculations presented in the original version of
this manuscript.
This research is supported by the US DOE under contract
DE-AC02-06CH11357.
    
\appendix
\section{Calculational details}
\label{appendix}
\subsection{Notation for spinor products}
\label{Notation}
We adopt the following notation for massless spinors,
\begin{eqnarray}
|i\rangle &= |i+\rangle = u_+(p_i), \; |i] &= |i-\rangle = u_-(p_i) \;, \nn \\
\langle i| &= \langle i-| = \bar{u}_-(p_i),\; [i| &= \langle i+| = \bar{u}_+(p_i)  \;.
\end{eqnarray}
Further the spinor products are defined as,
\begin{eqnarray}
\spa i.j &=& \langle i-|j+\rangle = \bar{u}_-(p_i) u_+(p_j) \;, \nn \\
\spb i.j &=& \langle i+|j-\rangle = \bar{u}_+(p_i) u_-(p_j) \;,
\end{eqnarray}
with $p_i,p_j$ massless particles. With our convention,
\begin{equation}
\spa i.j \; \spb j.i = 2 p_i \cdot p_j = s_{ij} \;.
\end{equation}

We shall use the 
standard trick~\cite{Kleiss:1985yh} of decomposing 
the massive momentum, $p^2=m_t^2$ into the sum of two massless momenta, $p = p^\flat + \alpha \eta$ with the constant
$\alpha$ given by, 
\beq
\alpha=\frac{m_t^2}{\langle \eta| \slsh{p}|\eta]} \;.
\eeq
We may write the massive spinors as combinations of massless spinors as follows,
\beqn
\bar{u}_{-}(p)&=&[\eta | (\slsh{p}+m_t) \frac{1}{[\eta \, p^\flat]},\;\;\;
\bar{u}_{+}(p)=\langle \eta_t | (\slsh{p}+m_t)  \frac{1}{\langle \eta_t \, p^\flat \rangle } \;, \\
v_{+}(p)&=&(\slsh{p}-m_t) |\eta\rangle \frac{1}{\langle p^\flat \, \eta\rangle},\;\;\;
v_{-}(p)=(\slsh{p}-m_t) |\eta] \frac{1}{[ p^\flat \, \eta]} \;.
\eeqn
The spin labels of the massless spinors $|\eta\rangle,|\eta ]$ 
encode the polarization information 
of the massive quarks and they are equivalent to helicities only in the massless limit.

\subsection{Lowest order matrix element}
We present results for the basic amplitude at leading order,
\beq \label{basicamp}
u (p_1) + b(p_2) \rightarrow l(p_3) + a(p_4) + t(p_5) + d(p_6) \;,
\eeq
where $l$, $a$ are the lepton and anti-lepton respectively and momentum labels for the particles
are given in parentheses.

We begin by introducing the relevant couplings that
appear in the calculation. The current for the emission of a $Z$ boson or virtual
photon that decays into a left-handed lepton pair enters with a strength,
\beq
V^L_j = Q_j q_e +L_j l_e s_{34} D_Z(s_{34}) \;, \qquad V^R_j = Q_j q_e +R_j l_e s_{34} D_Z(s_{34}) \;,
\eeq
where the superscript denotes the helicity of the outgoing quark and
the subscript the flavor of the quark from which the boson is emitted
($j=u,d$). In this formula the individual quark and lepton couplings
are themselves defined by,
\beqn
&& L_j =\frac{\tau_j-2 Q_j \sin^2 \theta_W }{\sin 2\theta_W } \;, \qquad R_j =\frac{ -2 Q_j \sin^2 \theta_W }{\sin 2\theta_W } \;, \\
&& l_e =\frac{-1-2 q_e  \sin^2 \theta_W }{\sin 2\theta_W } \;, \qquad r_e =\frac{ -2 q_e \sin^2 \theta_W }{\sin 2\theta_W }  \;,
\eeqn
where $q_e=-1$, $\tau_u = 1$ and $\tau_d = -1$. The $Z$ propagator denominator is,
\beq
D_Z(s_{34}) = \frac{1}{s_{34}-m_Z^2} \;.
\eeq

We first consider the case of a negative helicity outgoing lepton and a negative spin-label for the top quark. 
The contributions to the amplitudes, calculated in the Feynman gauge and labelled by the diagrams in Fig.~\ref{fig:STZ} are,
\begin{eqnarray}
&&  M^{(a,b)}(1^-_u,2^-_b,3^-_l,4^+_a,5^-_t,6^+_d) =
       D_W(s_{25})\frac{1}{s_{34}}  \nonumber  \\ 
&\times&     \left[\frac{V^L_u}{s_{134}} \spa{5^\flat}.{6} \spb1.4 \spab{3}.{1+4}.{2}
        -\frac{V^L_d}{s_{346}} \spa3.6 \spb1.2 \spab{5^\flat}.{3+6}.{4}\right]\\
&&  M^{(c,d)}(1^-_u,2^-_b,3^-_l,4^+_a,5^-_t,6^+_d) =
         \frac{D_W(s_{16})}{s_{34}} \Biggl[ -\frac{V^R_u  m_t^2}{(s_{345}-m_t^2)} 
           \frac{\spa{3}.{6} \spb{1}.{2} \spb{4}.{\eta}}{\spb{5^\flat}.{\eta}}
\nonumber \\
&+& \frac{V^L_d} {s_{234}} \spab{3}.{(2+4)}.{1} \spa{6}.{5^\flat} \spb{2}.{4}
                        - \frac{V^L_u}{(s_{345}-m_t^2)} \spab{6}.{(1+2)}.{4} \spa{3}.{5^\flat}\spb{1}.{2} 
          \Biggr]  \\
&&  M^{(e,f)}(1^-_u,2^-_b,3^-_l,4^+_a,5^-_t,6^+_d) =
       \frac{D_W(s_{25}) D_W(s_{16})}{s_{34}} \Biggl[ -\big(V^L_u-V^L_d\big) \nonumber \\ 
   &\times& \Big\{\spab{3}.{(1+6)}.{4} \spa{6}.{5^\flat} \spb{1}.{2} 
   +\spab{5^\flat}.{(1+6)}.{2} \spa{3}.{6} \spb{1}.{4}
   +\spab{6}.{(3+4)}.{1} \spa{3}.{5^\flat} \spb{2}.{4}
          \Big\} \nonumber \\
 &+& \frac{m_t^2}{2} \frac{\spa{3}.{6} \spb{1}.{4} \spb{2}.{\eta} }{\spb{5^\flat}.{\eta}}
\Big\{ V^L_u-V^L_d -V^R_u+V^R_d \Big\}  \Biggr] \\
&&  M^{(g)}(1^-_u,2^-_b,3^-_l,4^+_a,5^-_t,6^+_d) =
       \frac{D_W(s_{25}) D_W(s_{16})}{2 \sin^2 \theta_W s_{235}} 
  \left[  \spa{3}.{5^\flat} \spb{1}.{4} \spab{6}.{(1+4)}.{2}  \right]
\end{eqnarray}
For the case of a positive spin-label for the top quark we have,
\begin{eqnarray}
&&   M^{(a,b)}(1^-_u,2^-_b,3^-_l,4^+_a,5^+_t,6^+_d) =
       D_W(s_{25}) \frac{m_t}{s_{34}} \nonumber \\
&\times &      \Big[\frac{V^L_u}{s_{134}}   
          \frac{\spa{6}.{\eta} \spb1.4}{\spa{{5^\flat}}.{{\eta}}} \spab{3}.{(1+4)}.{2}
        + \frac{V^L_d}{s_{346}} 
          \frac{ \spa3.6  \spb1.2 }{\spa{{5^\flat}}.{{\eta}}} \spab{\eta}.{(3+6)}.4 \Big] \\
&&  M^{(c,d)}(1^-_u,2^-_b,3^-_l,4^+_a,5^+_t,6^+_d) =
         \frac{D_W(s_{16}) m_t }{s_{34}} \Big[ \frac{V^R_u}{(s_{345}-m_t^2)} \spa{3}.{6}\spb{1}.{2} \spb{4}.{5^\flat}  
\nonumber \\
&&      -\frac{V^L_d}{s_{234}} \frac{\spab{3}.{(2+4)}.{1} \spa{6}.{\eta}  \spb{2}.{4}}{\spa{5^\flat}.{\eta}}
        +\frac{V^L_u}{(s_{345}-m_t^2)}  \frac{\spab{6}.{(1+2)}.{4} \spa{3}.{\eta} \spb{1}.{2}} {\spa{5^\flat}.{\eta}} \Big] \\
&&  M^{(e,f)}(1^-_u,2^-_b,3^-_l,4^+_a,5^+_t,6^+_d) =
        D_W(s_{25}) D_W(s_{16})\frac{m_t}{s_{34}}  
       \Big[\big(V^L_u-V^L_d\big)\nonumber \\
   &\times &\Big\{ \frac{1}{\spa{5^\flat}.{\eta}}  (
          \spab{3}.{(1+6)}.{4} \spa{6}.{\eta} \spb{1}.{2}
          + \spab{\eta}.{(1+6)}.{2} \spa{3}.{6} \spb{1}.{4} 
          + \spab{6}.{(3+4)}.{1} \spa{3}.{\eta}  \spb{2}.{4} \Big\}  \nonumber \\
       &+& \frac{1}{2} \spa{3}.{6} \spb{1}.{4} \spb{2}.{5^\flat} 
\Big\{ V^L_u-V^L_d +V^R_u-V^R_d \Big\}\Big] \\
&& M^{(g)}(1^-_u,2^-_b,3^-_l,4^+_a,5^+_t,6^+_d) =
        -\frac{D_W(s_{25}) D_W(s_{16}) m_t }{2 \sin^2 \theta_W s_{235}}  \Big[
       \frac{\spa{3}.{\eta} \spb{1}.{4} \spab{6}.{(1+4)}.{2}}{\spa{5^\flat}.{\eta} }
          \Big]
\end{eqnarray}
Note that the opposite helicity combination for the lepton line is obtained by performing the flip
$ 3 \leftrightarrow 4, l_e \rightarrow r_e$ for $M^{(a,b)},M^{(c,d)},M^{(e,f)}$. 
The amplitude $M^{(g)}$ does not contribute for the opposite helicity.

The total leading order amplitude is obtained by summing these four subamplitudes.
In order to allow the $Z$ boson to be off-shell but still retain
gauge invariance, we use a simple prescription to incorporate the $Z$ width~\cite{Baur:1991pp}.
We use the propagator factor $D_Z(s_{34})$ in the amplitudes as written above
and then multiply the whole amplitude by,
\begin{equation}
\left( \frac{s_{34}-m_Z^2}{s_{34}-m_Z^2+im_Z\Gamma_Z} \right).
\end{equation}

\section{Associated production of a single top and Higgs boson}
\label{tHappendix}

\begin{figure}[t]
\begin{center}
\includegraphics[angle=270,width=10cm]{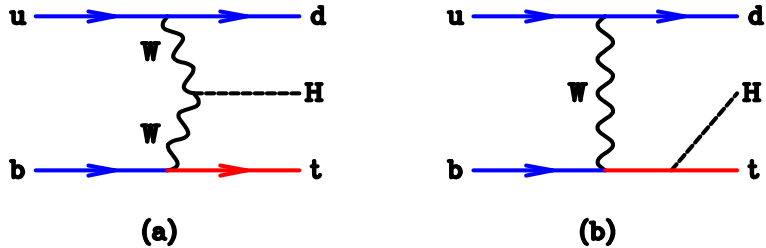}
\caption{Feynman graphs to calculate the lowest order amplitudes for single top + Higgs production.
The diagrams where the Higgs boson couples to the light quarks have been dropped.}
\label{fig:STH}
\end{center}
\end{figure}

In this appendix we briefly describe the NLO calculation of single top + Higgs boson production, which
is very similar in many respects to the single top~+~$Z$ process that is the main topic of this paper. 
In the limit in which the light quarks are taken to be massless, 
there are only two leading order diagrams, as shown in Figure~\ref{fig:STH},
with the Higgs boson attaching to either the top quark or the $t$-channel $W$ boson. 
This process has previously been considered in Refs.~\cite{Maltoni:2001hu,Barger:2009ky,Farina:2012xp}. The gauge cancellation between the two diagrams in Fig.~\ref{fig:STH} results in a smaller cross section compared to the associated pair production
mode, $t\bar t H$. In addition, because of the small branching ratios of a $126$~GeV Higgs boson to the cleanest modes ($H \to$ four leptons and
$H \to \gamma \gamma$), single top + $H$ production will be extremely challenging to observe. Nevertheless, like the $t\bar{t}H$ process,
this channel has the potential to measure the coupling of the
Higgs boson to the top quark. Reliable theoretical estimates for the $t\bar t H$ process, accurate to NLO,
are given in Refs.~\cite{Beenakker:2001rj,Reina:2001bc,Beenakker:2002nc,Dawson:2003zu} and including also
the effect of a parton shower in Refs.~\cite{Garzelli:2011vp,Frederix:2011zi}. Here we bring the accuracy of the
single top + $H$ channel to the NLO parton level.

Our results are calculated using the same numerical procedure described in Section~\ref{outline}. Due to the simplicity
of the scalar coupling of the Higgs, it is possible to immediately reduce the rank of the tensor integrals that appear
in the 1-loop calculation to a maximum of two. As a result we find that the calculation is significantly more stable
than the single top + $Z$ case, with an order of magnitude less events discarded due to insufficient numerical precision
in the pole terms (less than $0.005\%$). The renormalization of the Yukawa coupling of the Higgs boson to the top quark takes exactly the same
form as the renormalization of the $\varphi$ coupling already discussed in Section~\ref{outline}.

\begin{figure}[ht]
\begin{center}
\includegraphics[angle=0,width=0.45\textwidth]{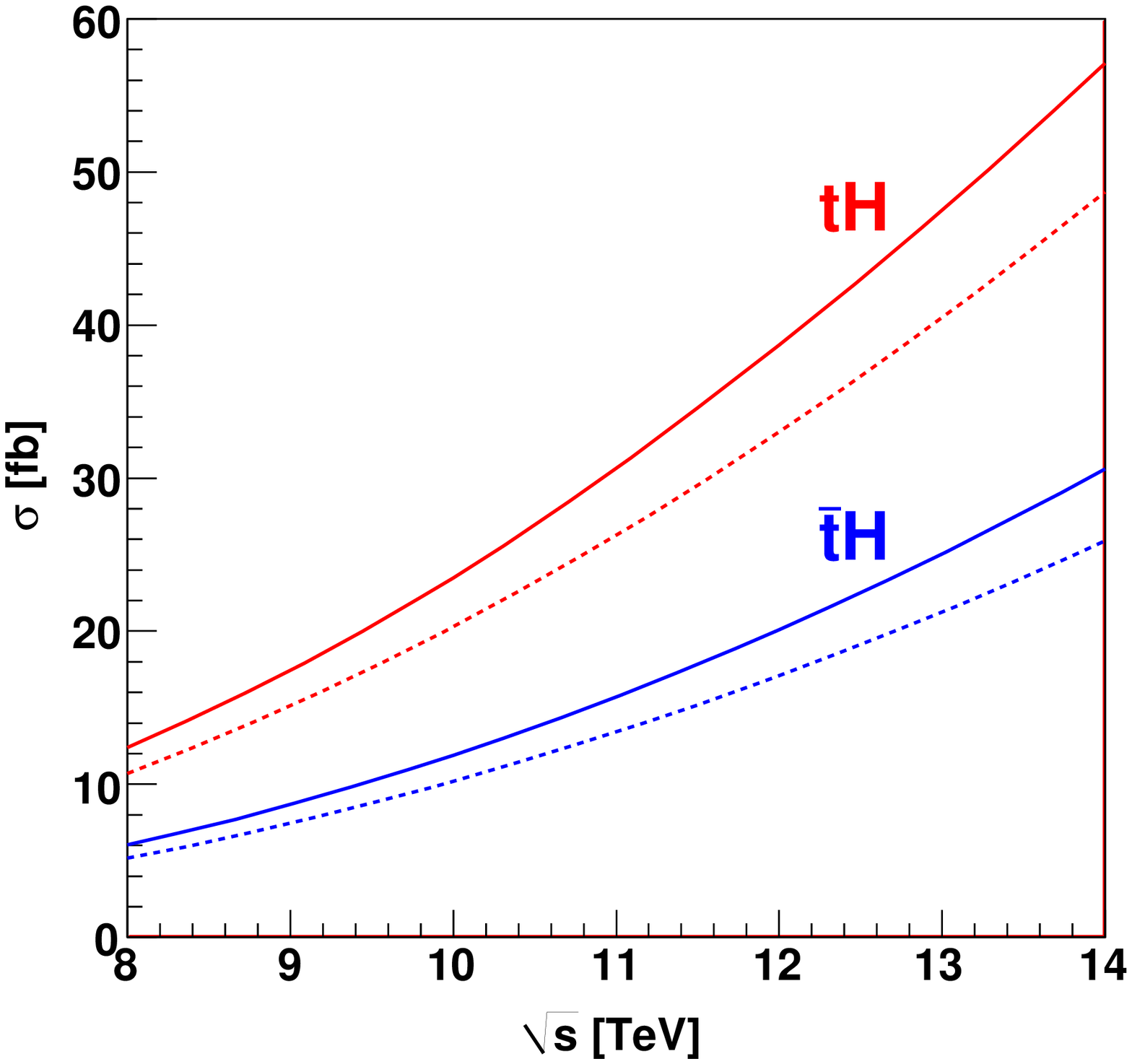}
\includegraphics[angle=0,width=0.45\textwidth]{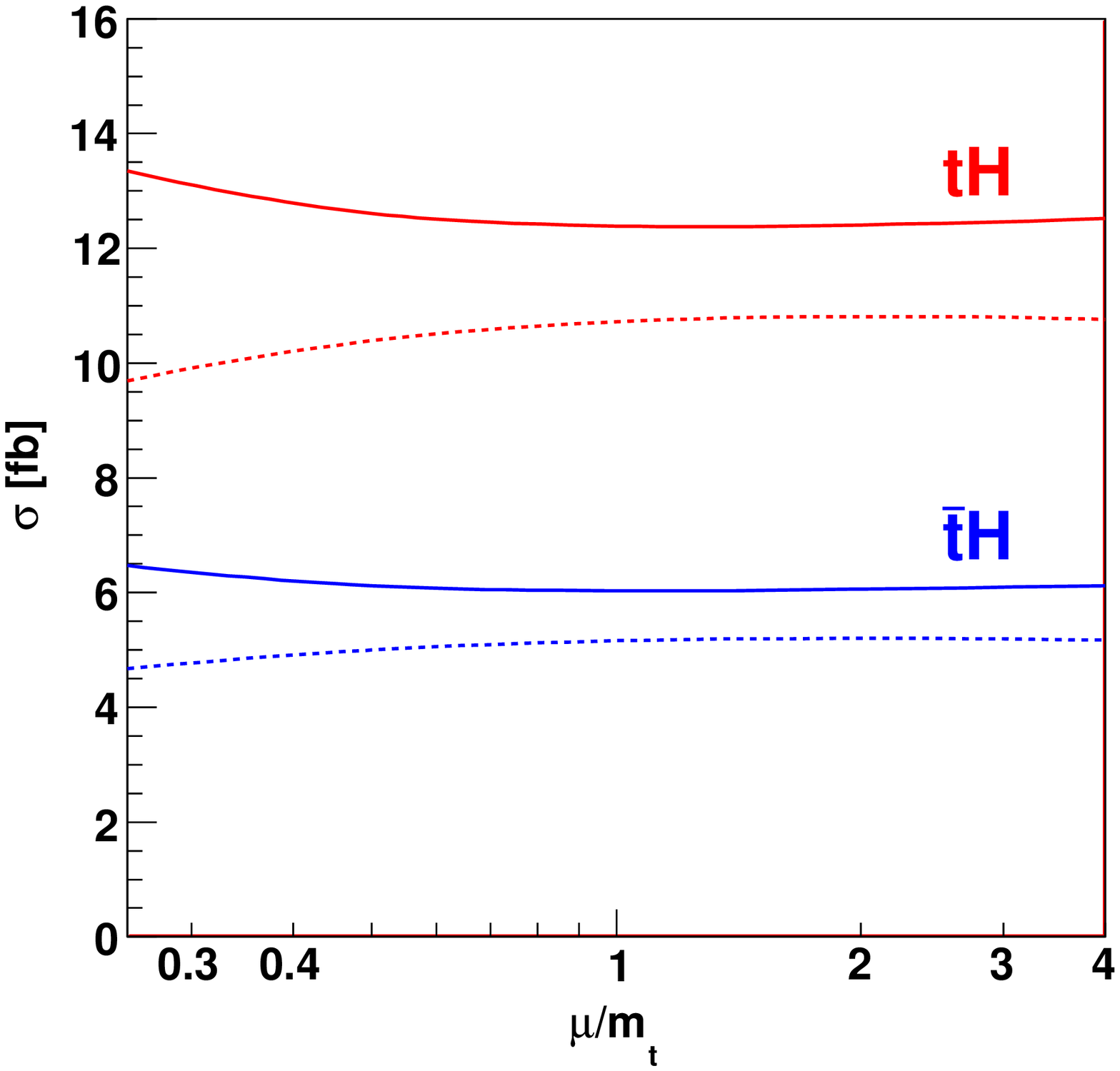}
\caption{Left: Cross sections for $tH$ and $\bar t H$ production as a function of $\sqrt s$.
Right: Scale dependence of single top + $H$ cross sections at $8$~TeV ($\mu = \mu_R = \mu_F$).
In both cases, the Higgs boson has a mass $m_H=126$~GeV and leading order predictions are shown as dashed lines,
next-to-leading order as solid lines.
}
\label{tHsqrts}
\end{center}
\end{figure}

For the results presented here we use $m_H=126$~GeV based on the first observation of a new boson at the LHC.
The cross sections for $tH$ and $\bar t H$ production as a function of the LHC operating
energy $\sqrt{s}$ are shown in Figure~\ref{tHsqrts} (left).
The effect of next-to-leading order corrections is larger than in the single top + $Z$ case,
with an increase in the cross section of approximately $15\%$ at NLO.
To investigate the scale dependence of this process we focus on the case $\sqrt s=8$~TeV.
In contrast to the production of single top~+~$Z$, in this case we find the largest scale dependence when
both renormalization and factorization scales are varied together. The results 
are shown in Figure~\ref{tHsqrts} (right), where we consider scale variation by a factor of four about the central value, $\mu=m_t$.
Once again the NLO scale dependence is very mild, as expected in an electroweak process.

This process has received considerable interest recently as a probe of non-standard couplings of the
Higgs boson to top quarks~\cite{Farina:2012xp,Biswas:2012bd}. If the couplings deviate
from their SM values (e.g. due to New Physics effects in loops) then the $tH$ cross-section
may be significantly enhanced. We allow the possibility of anomalous couplings in our code
to enable a NLO calculation of such effects.

\bibliography{top}
\bibliographystyle{JHEP}

\end{document}